
\hfuzz=8pt
\tolerance =10000
\magnification 1200
\catcode`@=11
\baselineskip=24pt
\nopagenumbers
{}~
\vskip 1truecm
{\centerline{\bf Criticality of lamellar surfaces by
conformational degrees of freedom}}
\vskip 1.5truecm
{\centerline {B.Bassetti, G.Mazzoletti}}
{\centerline { Universit\'a degli Studi di Milano, Dipartimento di Fisica}}
{\centerline { Via Celoria 16, 20123 Milano,Italia}}
\vskip 1truecm
{\centerline {P.Jona}}
{\centerline { Politecnico di Milano, Dipartimento di Fisica}}
{\centerline { Piazza Leonardo da Vinci 32, 20123 Milano,Italia}}
\vskip 2.5truecm
\noindent
{\it Shortened version of the title:}

{\bf " Criticality by conformational degrees of freedom"}

\noindent
{\it Selected PACS numbers:}

64.60~ - General studies on phase transitions

68.42~ - Surface phase transitions and critical phenomena

64.70M - Transitions in liquid crystals
\vfill\eject
{\bf ABSTRACT}
\vskip 1.5 truecm
A new model for lamellar surfaces formed by anisotropic
molecules is proposed.
The molecules  have internal degrees of freedom,
associated with their flexible section of length $N$
at zero temperature.
We obtain a 2D non-standard six vertex model, which is
exactly soluble and exhibits a finite order transition.
The order and the character of the transition are
determined by the dominant term in
the $1 \over N$-expansion of the interaction energy. The dependence of
the critical temperatures on $N$ is, instead, determined by the
non-leading terms in the same expansion.
\vfill\eject
{\bf 1. INTRODUCTION.}
\vskip 1.5 truecm
In this work we shall study the statistical properties
of a hamiltonian model for a
two dimensional object formed by flexible molecules
arranged in the ordinary 3D space.

\noindent
We think that this ideal model
not only can mimic  real physical {\it surfaces}
such as  layers of molecules
forming films and interfaces
or (semi)bilayers in biological membranes, but
also can be useful for the interpretation of
the layered structures, and related transitions,
observed in smectic phases of liquid crystals.

In view of our purposed description, two properties
of the elementary components are of main interest.
The first one is the presence in the molecule
of a (long) flexible section, or {\it tail}, typically an
alkylic sequence; the {\it conformational}
flexibility is associated with internal degrees
of freedom which represent the rotations around skeleton bonds
in the tail.
In agreement with the formalism of Volkenstein and Flory$^{(1,2)}$
who developed the conformational theory of chain molecules,
we shall refer in this paper
to {\it trans} planar chains and to chain {\it defects}
such as {\it kinks}, {\it antikinks} and their combinations;
we shall also  make use of Flory's definition of conformational
(intramolecular) energies $^{(2)}$.
The second structural property of interest is the presence
in the molecule of a rigid core, or {\it head},
carring an asymmetric charge distribution,
equivalent to a dipole;
the relative orientation of such a dipole with respect to
the molecular longitudinal axis is also meaningful in our analysis.
It is well known from experiments that systems
formed by molecules with these properties exhibit
a rich variety of mesophases (first classified by
Friedel$^{(3)}$) which, beyond the perfect
crystal, go toward the isotropic liquid trough a
progressive loss of intermolecular order$^{(4-10)}$.
Both thermotropic and lyotropic behaviours have
been observed.
In thermotropic systems, one has a pure substance
formed by mesogenic molecules; the
transitions between mesophases
occur by varying the temperature of the sample.
The order parameters describe the orientational
and the positional arrangement of molecules; they change
with temperature and are discontinuous at critical
temperatures.
In lyotropic systems, one has a solution where the
solute is a mesogenic molecule; the
phase transitions are achieved
by changing the concentration of the solute
or by changing the character (ionic strength)
of the solvent.

In this paper, we shall refer to thermotropic
compounds.  In this field, a wide collection of
experiments (see for instance ref.10),
investigated the relations between structural
properties of molecules and liquid crystalline
basic properties; the main interest has been
focussed on the  spatial organization of molecules
in the various mesophases, the behaviour of
critical temperatures in series of homologues
and the sequences of occurring mesophases.
In particular, the layered, {\it head to tail},
arrangement  of molecules was observed,
since 1975, by Leadbetter {\it et al.}$^{(11,12)}$
in  the smectic A phase.
By X-ray diffraction on cyanobiphenyls
compounds, they confirmed that the layer
spacing is greater than the molecular length,
consistently with the proposed setting
of pairs of molecules, forming dimers, with
heads facing one an other in the center of the layer
and tails pointing outward in opposite directions.
Analogous layered arrangements were observed
in smectic phases for many other thermotropic
compounds; as examples we may quote the
nitroderivatives$^{(13)}$ and
cyanatophenylbenzoates$^{(14)}$ with alkylic tails.
In these substances many different $S_A$ phases
are recognized by X-ray  diffraction, calorimetric$^{(13)}$
and dielectric$^{(14)}$ measures; in each phase
the head to tail setting occurs with a
different extent of molecular overlapping and
layer thickness.
A great variety of Smectic- Smectic transitions
may occur; in particular, the transition
$S_{A1} \rightarrow  S_{A2}$,
from the phase with layer thickness equal to
the molecular length,
to the phase with layer thickness twice
the molecular length$^{(13)}$ is observed
for increasing temperature; the transition $S_{A1}$
to nematic is also observed$^{(13,14)}$.

\noindent
Two notes are in order about these facts.
First, the persistence of layered structures in
the different smectic phases suggests
the interaction between adjacent layers to be weak,
compared with the {\it intralayer} forces;
for this reason we think that it is meningful
to study the single, smectic-like layer
as a two dimensional, independent, neutral and
thermotropic object in the 3D space.
Second, the {\it head to tail} setting of adjacent
molecules suggests the antiferroelectic arrangement of
molecular dipoles,  parallel to the longitudinal
molecular axis,  to be compatible with the alignment
of molecular axis.

{}From a theoretical point of view, the formation of
mesophases is equivalent to a sequence of symmetry
breakings.
Starting from the completely symmetric 3D phase
(isotropic liquid), one expects the strongly asymmetric
interactions between elementary constituents to promote,
step by step, the orientational order, the
formation of stable and weakly interacting objects
with lower dimension than that of the environment
(such as 2D smectic layers, or 1D columnar structures),
up to the perfect crystal$^{(15)}$.
However, this progam is rather difficult to execute.

\noindent
The first step has been overcome by Onsager,
since 1949$^{(16,17)}$, with an exactly soluble model
for a system of rigid rods with
first order isotropic to nematic (I-N) transition.
There is a wide and rich literature
about this transition; one has the phenomenological
description by Mayer Saupe$^{(18)}$,
the Landau theory of orientational order parameter
by De Gennes  $^{(19)}$, and the order parameters
by McMillan's model$^{(20,21)}$.
The presence of a flexible section in the molecule,
and its effects in the I-N transition, was
considered in Marcelja's self consistent analysis
of chain ordering in liquid crystals$^{(23)}$
and in Pink's model of odd-even effect$^{(24)}$;
the problem was also investigated by
Martire$^{(25)}$ and Luckhurst$^{(26)}$ with
molecular field methods.
The orientational transition
was also studied by Monte Carlo simulations;
we recall the first calculation by Lebwohl
and Lasher $^{(27)}$ and, more recently,
the planar versions of the $RP^n$ model$^{(28-30)}$.

For other transitions, between different smectic
phases (S-S) or for smectic to nematic
transition (S-N), the research is not so
widely developed; in particular,
there are no exactly solved models and
one turns to phenomenological theories
of order parameters or to numerical simulations.
For liquid crystals, the translational order
parameter for S-N transition was defined
by McMillan$^{(20,21)}$ and compared
with the phenomenology of smectic
phases$^{(22)}$. The problem about the
role played by conformational flexibility
in these phases and related transitions is
avoided: the molecules are ever considered
as rigid rods.

\noindent
On the other hand, in the field of biological
membranes, Nagle$^{(31)}$ and,
independently, Marcelja $^{(32)}$
developed two models for the transitional
behaviour of molecular chains in smectic-like
phases of lipid mono and bilayers.
In these models, the attention is pointed
on the conformational collapse of the
constituents rather than on the effect
of such a collapse on the surface of the
biomembrane.

Before introducing the model, which is
addressed to the study  of $S-S$ and
$S-N$ transitions, we note
that the criticality of layered structures
is of main interest in a rather different
context, namely in studying the interface
formation between (e.g) two solids or
film deposition; in this case the geometry
of the surface is characterized by the formation
of ledges and terraces, with the typical
{\it roughening} transition. Surface structures and
related phase transitions have been
studied with exactly soluble models$^{(33)}$;
we recall the ones by Van Beijeren
(BC-Solid-On-Solid $^{(34,35)}$)
and by Burton (Terrace-Ledge-Kink $^{(36)}$),
firstly introduced for the crystal-vacuum
interface. The common feature of
these models is  that they can
implement a conservation law
by a {\it six vertex} lattice system,
with exactly known $^{(37-39)}$ solution.
Monte Carlo calculations on specific
realizations of BCSOS or TLK models
simulate (e.g) adatoms forming thin
films $^{(40,41)}$ or interfacial
molecules $^{(42)}$.

In this work, we present an exactly soluble
model for a {\it single} molecular surface,
namely for a 2D object like a {\it free}
smectic layer of a thermotropic liquid crystal.
Together with the general structural properties
previously required for the constituent
molecules (the presence of one head and one tail),
we consider the following distinctive features:

a) -  we are interested in a system whose
{\it intra}molecular energies, associated with conformational
chain defects, are of the same order of magnitude as the
{\it inter}molecular energies;

b) - we represent the charge distribution
associated with the molecular head by a dipole
{\it parallel} to the  longitudinal
axis of the molecule;

c) - we assume an antiferromagnetic structure of the
foundamental state; the constituents are aligned
with an {\it head to tail} setting.

\noindent
There are at least two physical reasons for assuming
the property a).
First, in many molecular systems with smectic transitions,
the transition heat is of the same order as the melting
heat of the crystal formed by the tails only
(typically n-alkanes lamellae)$^{(4)}$.
Second, in the same systems, IR and Raman spectra
directly show that the alkylic part of the molecular
layer is liquid like  at the critical temperature;
in fact, the spectroscopic markers of the conformational
defects characterizing the liquid phase of n-alkanes
are clearly observed$^{(43)}$.

\noindent
As for the requests in b) and c), we stress that
they are not independent one each other.
In fact, constituents with head dipoles
orthogonal to the longitudinal axis, should
be compatible with a layered structure in which
molecules have an {\it head to head} setting
(like in a semi-bilayer, in which the alkylic
chains point in the same direction).
In this case we expect to be addressed to
the phenomenology of lyotropic systems,
in which the transitional properties are
determined by the dominant interaction between
polar heads and solvent;
alternaltively, for thermotropic compounds,
we expect the system to have a transitional
behaviour dominated by vortex formation
in the plane of molecular heads$^{(44)}$.

\noindent
Assuming longitudinal head dipoles instead, the
arrangement of molecules, in a temperature regime
with high orientational order, is compatible with the
head to tail setting.
We think that, in this case, there is an effective
competition between the dipole-dipole and the pure chain-chain
contributions to interaction energy;
pairs of opposite dipoles will have a tendency to
reach an antiparallel setting with aligned centers;
on the other hand tails will prefer to attain the full
extended trans planar conformation with aligned
centers of mass.
In presence of a rigid core, this
competition leads to an interaction potential
with (at least) two different minima which, roughly speaking,
correspond to the maximum coupling of dipoles and chains
respectively.
Let us consider, in the 3D space, the packing
of pair of molecules
in one of the equilibrium relative
positions; we shall obtain the layered
smectic-like structure, namely the single surface
of interest. Furthermore, different geometries
select different classes of conformational defects,
which are compatible with alternative packings.
We can study the stability, the thickness
and the degree of conformational disorder
of such a surface as a function of
both the conformational energy
and the energy difference between the minima
of the two-body potential.

In section 2. we construct the {\it bare model}
of the surface, obtained neglecting the conformational degrees
of freedom. We show that it belongs to the class of
{\it vertex models} $^{(38)}$.
The specific character
of the interaction leads to a definition of
vertex states and related energies, which are different from those
in the vertex models studied in the literature. The symmetries and the
conservations typical of the solved classical model do not apply
in our case, except for a special choice of the interaction
constants.  We show that, in the general case, the bare model
can't have thermotropic phase transitions.

In section 3. we take into account the conformational degrees
of freedom and we show that the vertex states have an internal
degeneracy. Therefore the model may exhibit
a thermotropic transition of
{\it finite} order. The thickness of the surface at
critical point is also studied.

In section 4. we analyze the critical
dependence of the energy parameters
of the model on the length of the molecule flexible section.
We obtain the dependence of the order of the transition and of
the critical temperature on such length.
We then compare our results
with some experimental data.

In the last section we point out two lines
of developement of our
model: the first in view of a description
of bilayers; the second
toward an interpretation of the multiplicity of mesophases
in liquid crystals.

\vskip 3truecm
{\bf 2. THE BARE MODEL}
\vskip 1.5truecm

We are interested in modelling a surface exhibiting
orientationally ordered phases, so that
we will assume the existence of a fixed director
along which the molecules are aligned.

At first, we don't consider the flexibility
of molecular tails: {\it in the bare model the
conformational degrees of freedom are frozen}, the
molecules are rigid objects.

Representing the surface through the centers of mass
of the molecules, we consider their projections
in the orthogonal plane to the director;
we assume that
such projections define a square $L \times L$ lattice,
formed by two intertwined sublattices
$A$ and $B$. Each lattice site in $A$ ($B$) is
occupied by a molecule with head dipole
oriented up (down) with respect to the
lattice plane; in this way we realize the
head-to-tail setting of molecules, which,
as described in the introduction, has been
observed in smectic layers$^{(11-14)}$.

The translational degrees of freedom of
the molecules are constrained along the
director; moreover, we don't consider
vibrations (relative oscillations along the
director, around the configurations which
minimize the two-body energy).

\noindent
The surface can now be characterized
by  the scalar field  $\{h_{I,J}\}$
of the {\it altimetric heights}
of the molecular centers of mass,
relative to a given reference.

For a pair of molecules corresponding
to adjacent sites in the lattice,
the competition between dipole-dipole
and chain-chain interactions
determines two stable {\it link states}
(with minimal energy). If the attractive
chain-chain interaction prevails,
we have two neighbouring molecules
with tails facing one another
and head dipoles forced in a non
equilibrium (relative) position
(see fig. 1a); this configuration
defines the link state (0).
If, instead, the antiferromagnetic
interaction between head dipoles prevails,
we have the link state (1), representing
facing heads and tails forced to point toward
opposite directions (see fig. 1a).
Note that the direction of the line containing
the centers of mass (local surface orientation)
with respect to the director is different in the
two configurations.

\noindent
Both link states (0) and (1) are minima
for the pair interaction.
In fact, a displacement around the configuration 1,
increases the dipole-dipole contribution to
interaction energy, whereas the tails are
too far to compensate.
Analogously, a displacement around 0,
decreases the attraction between the tails,
but the opposite dipoles are again too far.
The change from link-state (0) to (1) is obtained
by shifting an $A$ site molecule downward,
or by shifting a $B$ site molecule
upward (fig. 1b); since vibrations are excluded,
the single molecule may only shift up or down
of one unit length, relatively to its neighbours.

In the bare model, given a distribution of
links on the lattice, the height of a molecule
in the site $(I,J)$, relative to the height of
the molecule in the site $(0,0)$ (of type A),
is a function of the values ($0$ and $1$)
of link states. We have:

$$h_{I,J}=\sum_{j=0}^{J-1}
(-)^j x_{0,j} +(-)^{J} \sum_{i=0}^{I-1} (-)^i
y_{i,J}
\eqno (1)$$

\noindent
where $x_{i,j},y_{i,j}$ are the values
of the links $(i,j),(i,j+1)$
and $(i,j),(i+1,j)$ respectively.

\noindent
Note that the sign of the height variation
between a site and its nearest neighbour depends
on the parity of the site; furthermore,
the sum of the height variations, over any loop
on lattice sites,  must be zero.
This is a constraint for the
distributions of links on the lattice.

Let us define a {\it plaquette} as a loop of
four links and a {\it plaquette state} as the
set of values of such links, the constraint
selects  6 of the 16 possible plaquette states,
namely those where:
$$ x_{I,J}-y_{I,J+1}-x_{I+1,J}+y_{I,J}=0 \eqno (2)$$

The previous description can be given in the
formalism of {\it vertex models} $^{(38)}$.

\noindent
The plaquettes are identified with the sites of the dual lattice
and the plaquette states with the vertices on such sites.
A {\it vertex} is defined by four arrows
entering or exiting the site
(arrows-in/out) and connecting it with the nearest neighbours;
it is easy to set a one-to-one correspondence between plaquette
states and vertices.

\noindent
The so-called 8 vertex (8V) model is built on
vertices with an even number of arrows-in
(8 allowed states per site); the 6 vertex (6V)
models have 6 allowed states per site. In particular
the vertices of the {\it classical} 6V model must
satisfy the so-called {\it ice rule}:
two arrows-in  and two arrows-out.

\noindent
The configurations of the model are sets of
compatible nearest neighbouring
vertices; the energy of a configuration is
the sum of the energies assigned
to the vertices.  In the classical, exactly
soluble models, the energy is
invariant with arrows reversing;
if such symmetry is broken, one has
a model with {\it external field}.

The allowed plaquette states of our model are mapped in 6 of the 8
vertices of the 8V model, as represented in fig.2.; this procedure
formally leads to a 6V model, but,
in comparison to the classical 6V, we have a
different choice among the 8V vertices (vertices 7 and
8 instead of 3 and 4).

\noindent
The configuration energy in the bare
model is the sum of plaquette energies;
for symmetry reasons
the plaquette states 5,6,7,8 must have the same energy
($\epsilon $), whereas the energy of states 1 and 2 can be
set to the values $-\omega/2$ and $+\omega/2$
respectively, by an appropriate choice of the zero for
the energy scale.

\noindent
In the special case $\omega=0$ (vertices 1 and 2 with
the same energy),
we can use the symmetry properties of the
partition function of the 8V model with zero field
(ref 38.ch.10) to prove that {\it the partition function of
the bare model is equal to the partition function of the
classical 6V model with ice rule}$^{(45)}$

\noindent
Therefore, in this special case, our bare model is equivalent to
the 6V model and it can be exactly solved.

\noindent
But, there is no physical
reason to assume that tail-tail interactions
have the same energy as dipole-dipole interactions,
so that, in general, vertices 1 and 2 have {\it unequal}
energies. In this case, the symmetry property, which is crucial in
proving the above statement, cannot be ge\-ne\-ra\-li\-zed;
it follows that, in a physically meaningful regime, our
bare model is not equivalent to the classical 6V.
Indeed, we have a specific version of the 8V model
{\it with external field}  ($\omega $);
even if, to our knowledge$^{(33,38)}$,
the 8V model has not been solved exactly
with non vanishing field, the specific features
of the present model make it easy to study.

\noindent
{}From the convexity properties of the free energy
and of the order parameter as functions of the
field $\omega$, we can prove that
{\it in the bare model the free energy
is differentiable with
respect to $\omega$ except, at most, in
$\omega =0$}.$^{(45)}$

\noindent
It follows that we have no chance for
thermotropic transitions
in the fully oriented physical regime,
as far as rigid molecules are considered.

In the next section we shall show that the special case
$\omega=0$ has a physical meaning if the
internal degrees of freedom of the flexible molecules
are considered.

\noindent
For this reason, we recall here the results of
the classical 6V model
that are relevant to our description.

\noindent
{\item - According to the exact solution$^{(38)}$, one has two
possibilities.
First: if the energy $\epsilon$ is strictly negative,
the system is disordered; at any temperature the free
energy is analytic and the path of
the point representing the system
in phase space lies in the disordered phase region.
Second: if $\epsilon >0$, for low temperatures the system
is in the antiferroelectric phase and has a phase transition,
of infinite order, at the critical temperature
$\beta_C={{ln(2)}\over {\epsilon}}$, to the
disordered phase. }

\noindent
{\item - The classical definition of order parameter,
becomes, in  our case:
$$m= -{ {\partial }\over
{\partial \omega}} \ln Z |_{\omega=0}=
\langle {1\over L^2} [n_1-n_2]\rangle \eqno (3)$$
namely  the expectation of the difference
between the numbers of
plaquettes of type 1 ($n_1$) and 2 ($n_2$). }

\noindent
{\item - The correlation length $\xi $,
which is obtained from the
behaviour of ${{\partial ^2 \ln Z}\over
{\partial \omega_{I,J} \partial \omega_{I',J'}}}$
for large distance of the sites $(I,J)$ and $(I',J')$,
is finite for $\beta > \beta_C$
and diverges for $\beta < \beta_C$ (ref 38.ch.8). }

\noindent
{\item - The susceptibility $\chi$ associated with
$m$, defined as
$\chi= {{\partial ^2}\over {\partial \omega^2}}\ln Z$ ,
turns out to be  finite for $\beta > \beta_C$
and diverges for $\beta < \beta_C$ $^{(35,39)}$. }

In order to refer our bare model to known results,
the altimetric description suggests the comparison
with the BCSOS model $^{(34,33)}$.
Therefore, we represent each plaquette by the
mean height of the four
constituent molecules and we identify
the order parameter $m$ in eq. (3)
with the {\it roughness} of the surface.

\noindent
We carried on a numerical calculation,
by transfer matrix method,
for the bare model with $\omega =0 $; in fact,
in terms of our link variables the transfer
matrix has an iterative structure which is very
simple and easily computable $^{(45)}$.
In particular, we checked the critical
behaviour of the order parameter $m$ and of
the associated susceptibility $\chi$.

\noindent
Our results about the order parameter, reported in
fig.3, approach, at the critical temperature $T_C$,
the infinite system asymptotic form, predicted
by the exact solution
($m\sim (T_C-T)^{-1/2}\exp \{-A(T_C-T)^{-1/2}\}$).
The associated susceptibility
(fig.4a) and its critical divergence with lattice size
($\chi_L(T_C) \sim L$, see fig.4b)
show a good agreement with
Montecarlo calculations on BCSOS$^{(40)}$.

\noindent
With reference to fig.3, we note that the
order parameter, as a function of temperature, is
{\it antisymmetric} with $\omega \rightarrow 0_{\pm} $
(we shall indicate these functions with
$m^+(\beta)=-m^-(\beta)$).
This fact, which comes from the invariance of the partition
function of the bare model with respect to the exchange
$ \omega \leftrightarrow -\omega$, will
be crucial in the following.

\noindent
We are also interested in evaluating the
dispersion $\delta^2( h)$ of the height,
related to the {\it thickness} of the
surface$^{(40,41)}$.
According to the BCSOS model, $\delta^2( h)$
is expected to behave as $\ln \xi$ for $\beta > \beta_C$
and to diverge as $\ln L$ with lattice size
$L$ for $\beta < \beta_C$ ~$^{(35,41)}$.
{}From  a phenomenological point of view,
we shall associate such a divergence with
the disgregation of the surface (see sec.3)

\vskip 3truecm
{\bf 3. SWITCHING ON THE INTERNAL DEGREES OF FREEDOM}
 \vskip 1.5 truecm
In this section, we shall consider the conformational
degrees of freedom and prove that their
switching on gives rise to a thermotropic
transition of finite order.

According to Flory's model of free chains,
the single molecular tail is described in terms of
conformational states. As we did for the case of alkanes
lamellae$^{(46)}$, we use a collective conformational
variable $\Delta$ and write the {\it chain partition function}
as a hierarchical sum:
$$Z_{Tot}=1+Z_{\Delta=1}+Z_{\Delta=2}+\cdots\eqno (4)$$
\noindent
Each term in the sum corresponds to a class of states
characterized by conformations with a fixed {\it transversal width}
$\Delta$ (with respect to the director).
The first term (1) is the contribution of the {\it all trans} chains;
the second term (with $\Delta=1$) comes
from {\it1-kink} or {\it 1-antikink} conformations and from all
conformations with sequences of alternating {\it kink} and
{\it  antikink} ; the $\Delta=2$ term
collects all the contributions from conformations with twice
the unitary width, and so on.

To construct the partition function of the system,
we have to define the structure of the interaction between molecules
which have different tail conformations and belong to different
plaquette configurations.

\noindent
We assume  that the multiplicity of available
conformational states depends, for each molecule, on the type of
plaquette the molecule belongs to; the intermolecular
interaction energies, instead, are assumed to be
independent of the particular
conformational state (provided it is available).
This is argumented as follows.
Let us consider a {\it dressed} plaquette of type 1
(see fig. 2; the molecules have facing tails and may
form conformational defects); since the tails are strictly
packed, they can only contain defects
with small width $\Delta $ and the
interaction is independent of the defect site
in the chains.
Considering now a dressed plaquette of type 2 (facing heads),
the molecular tails are rather free and then
may contain conformational defects with large $\Delta$.

\noindent
{}From a general point of view, these assumptions represent
a {\it packing-flexibility} constraint whose
specific realization (see sec.4) will not
affect the general results discussed in this section.

We write the {\it partition function of the system} in the form:
$$Z=\sum_{\{P_{I}\}}e^{-\beta\sum_{I}h(P_{I})}\sum_{\{S_{V|P}\}}
e^{-\beta\sum_{V}h(S_{V})}\eqno(5)$$

\noindent
where the first sum is over the plaquette states $\{P_I\}$
of allowed configurations and the second sum is over the
conformational states of the molecule in the lattice sites $V$,
constrained by the packing-flexibility condition with respect to
the four adjacent plaquettes.
The quantity $h(S_V)$ is the Flory energy of the state $S_V$ and
$h(P)$ is the plaquette energy.

\noindent
We can factorize the packing-flexibility constraint
in such a way that we can sum over the conformational
states and, using the geometry of
the lattice, share the result among the plaquettes.
The total contribution to the Z function
from a single dressed plaquette takes the form:
$$e^{-\beta h(P)}[Z_{[1]}\delta_{1,P}+
Z_{[2]}\delta_{2,P}+Z_{[0]}
\sum_{\chi=5}^{8}\delta_{\chi,P}]\eqno(6)$$

\noindent
where  $Z_{[1]}$,$Z_{[2]}$ are the partition functions
of a single chain with states compa\-tible with the plaquette
of type 1 and 2 respectively and $Z_{[0]}$ refers to
the other configurations.

\noindent
In terms of the plaquette energies $\pm{{\omega }\over {2}}$
(states 1 and 2 ) and $\epsilon$ (states 5,6,7,8) we have:
$$Z(\beta,\epsilon,\omega)=
\sum_{\{P_{I}\}}
e^{-\beta\sum_{I}
[{-\omega\over 2}-{1\over\beta}\ln Z_{[1]}(\beta)]
\delta_{1,P_{I}}}\cdot$$
$$\cdot e^{-\beta\sum_{I}[{\omega\over 2}-
{1\over\beta}\ln Z_{[2]}(\beta)]
\delta_{2,P_{I}}}\cdot$$
$$\cdot e^{-\beta\sum_{I} [\epsilon-{1\over \beta}
\ln Z_{[0]}(\beta)]\sum_{\chi=5}^{8}
\delta_{\chi,P_{I}}}
\eqno(7)$$

\noindent
One easily sees that this partition function has
the same form as in the bare model:
due to the conformational degrees
of freedom, the bare interaction constants become
the effective, temperature dependent, energy parameters:

$$\overline{\omega}=
\omega+{1\over \beta} \ln {Z_{[1]}\over Z_{[2]}} \eqno(8a)$$

$$\overline{\epsilon}=\epsilon-{1\over \beta}
\ln [{Z_{[0]}\over Z_{[2]}}
({Z_{[1]}\over Z_{[2]}})^{1/2}]\eqno(8b)$$

\noindent
The quantity ${{Z_{[1]}}\over {Z_{[2]}}}$
is a function of both the chain
length $N$ and $\beta$; with fixed $N$,
it can be written as a ratio of polynomials $^{(46)}$;
for $\beta\to\infty$ such ratio tends
to 1  and for $\beta\to 0$ it tends to the ratio between the
number of constrained states (small $\Delta $) and the number of
free states (large $\Delta $).

\noindent
The physical meaning of the additional terms
in expressions (8a,b) -~for brief
$f_{[1]}^{(N)}$ and  $f_{[0]}^{(N)}$~- is
the  difference between the free energy of a
chain without and with steric constraints.

\noindent
The function $f_{[1]}^{(N)}(\beta)$ is negative, monotone,
continuously differentiable with respect to $\beta$;
{}~$f_{[0]}^{(N)}(\beta)$ has analogous  properties.

{}From what precedes it follows that, with fixed $N$ and
provided $Z_{[1]}\not = Z_{[2]}$, one
inverse temperature $\beta^*$ {\it may exist}, such that:

$$f_{[1]}^{(N)}(\beta^{*})=-\omega\eqno(9)$$

For any $\beta\not=\beta^{*}$, since the energy
$\overline{\omega}$ is not equal to zero, with
the analysis in section~2., we {\it exclude} the occurrence
of {\it any} phase transition.

If $\beta=\beta^{*}$, the plaquettes of type
$1$ and  $2$ have zero effective energy,
whereas the others have positive effective energy
$\overline{\epsilon}(\beta^{*})$. It follows that,
at $\beta=\beta^*$, the equivalence condition with
the classical 6V model is verified.

\noindent
In order to study the behaviour of the free energy
of the system $f(\beta,\overline{\epsilon},\overline{\omega})$
as $\beta \to \beta^*_\pm$,
we consider the derivative of
$f(\beta,\overline{\epsilon},\overline{\omega})$
with respect to $\beta $. We have:
$$({d f\over d\beta})=({\partial f\over
\partial\beta})+({\partial f\over
\partial\overline{\omega}})({\partial
\overline{\omega}\over \partial\beta})+({\partial f\over
\partial\overline{\epsilon}})
({\partial\overline{\epsilon}\over
\partial\beta})\eqno(10)$$

\noindent
As $\beta \to \beta^{*}_{\pm}$, the partial derivatives
${\partial f\over
\partial\overline{\omega}}$ and  ${\partial f\over
\partial\overline{\epsilon}}$ approach the values of the order
parameters, calculated at $\beta^*$,
of the bare model with vanishing
external field ($\omega \to 0_{\pm}$);
in particular, due to the antisymmetry of $m$, we can write:
$$({\partial f\over
\partial\overline{\omega}})_{\beta_{\pm}^{*}}=m^{\pm}(\beta^{*})
\eqno(11)$$

\noindent
Now we compare the solution $\beta^*$ of eq.(9) with
the critical inverse temperature $\beta_C$ of the
classical 6V model.

\noindent
If $\beta^{*}>\beta_{C}$, we have:
$$m^{+}(\beta^{*})=-m^{-}(\beta^{*})\not =0\eqno(12)$$
\noindent
namely, the first derivative of the free energy is
{\it discontinuous}.

\noindent
In this case $\beta^*$ {\it is the critical
inverse temperature of a first order thermotropic transition}.

\noindent
The {\it latent heat} at the transition $\Delta H$ is the
difference between
the right and the left derivatives calculated at $\beta^*$
in eq. (10); we have easily:
$$\Delta H=2m(\beta^{*})[\beta^* {\partial f_{[1]}^{(N)}\over
\partial\beta}|_{\beta=\beta^{*}}]\eqno(13)$$

\noindent
In the above expression, the quantity
$\beta^*{\partial f_{[1]}^{(N)}\over\partial\beta}$
represents the {\it melting heat} of the tails; in fact,
$f_{[1]}^{(N)}$ is the difference between the
free energy of tails with straight conformations
and with completely disordered conformations.
Since the order of magnitude of $m$ is $1$,
the transitional latent heat and the melting
heat of the tails are really comparable.

\noindent
It is interesting to consider the critical behaviour of
the following variance, which, in the limit of infinite size,
represents the {\it thickness} of the surface:
$$\delta^2(s)= \delta^2(h)+\delta^2(D)\eqno (14)$$

\noindent
In this expression, the quantity $\delta^2(h)$ is the
dispersion of the height of the molecular centers of mass,
as defined in the bare model; the quantity $\delta^2 (D)$ is
the dispersion of the end-to-end distance of the molecular tails.
At $\beta^*$, the dispersion $\delta^2(h)$ is finite
and continuous (in fact, from the classical results,
since $\beta^*>\beta_C$,
one has $\delta^2(h) \sim \ln \xi$ with $\xi$ finite).
On the other hand, since the molecular tails undergo
conformational transition at $\beta^*$
(associated with the change from plaquette of type 1
to plaquette of type 2), the quantity
$\delta^2(D)$ abruptly changes from the
characteristic value of straight conformations to
that of disordered chains: both such values are finite.
This proves that, at the first order transition, the
thickness of the surface is {\it discontinuous}, but
{\it finite}.

\noindent
Last, we recall that the {\it orientation} of the
longitudinal axis of the molecules with respect to the surface
plane changes in going from link (0) to link (1); as a
consequence the mean orientation of the
layered molecules is {\it discontinuous} at the transition.

\noindent
In conclusion, from a phenomenological point of view, the
first order thermotropic transition changes the (mean)
geometry and the mutual positions of the molecules,
but {\it preserves the surface}.
This behaviour has been observed in
smectic-smectic transitions; as said in the
introduction, experiments show that
molecules maintain the layered arrangement
in the smectic phases, even if with different
layer thickness and molecular overlapping
(see for instance the $S_{A1} - S_{A2}$ transition
described in ref.13); it is also shown that
the orientation of molecular axis with respect
to the layer is different in the various
smectic phases $^{(9,10)}$.

We now consider the case $\beta^{*}<\beta_{C}$.

\noindent
One has:
$$m^{+}(\beta^{*})=m^{-}(\beta^{*})=0\eqno(15)$$
which implies that the first derivative of the free energy
is {\it continuous} at $\beta^*$; in this case no first order
transition occurs.

\noindent
Clearly, the bare model critical behaviours are again
maintained around $\beta^*$; in particular, the
susceptibility $\chi$ diverges.
If we consider the specific heat $c_v$, at $\beta^*$
we expect the same divergence
as $\chi$ (in fact, $c_v\sim {{\partial^2 f}
\over {\partial \beta^2}}\chi$).
This behaviour suggests the existence of a {\it second order}
transition.
{}From what said in sec.2, the fact that the zero field
condition $\overline \omega =0$ is reached at $\beta ^* < \beta_C$
implies that the correlation length diverges at the critical point.
In this situation, also the variance $\delta^2(s)$ {\it diverges},
due to the size divergence of $\delta^2(h)$.

\noindent
We associate this transition with the {\it disgregation}
of the surface, in fact the divergence
of $\delta^2(s)$  means that the long range correlation
between heights at different sites is lost.
With reference to the phenomenology of
nematics$^{(9,10)}$, we note that our model
behaves like a smectic layer of a system
undergoing S-N transition. In fact, we are
in a fully oriented regime
and we go from a phase with strong long
range correlation of heigths (the layered structure
of smectics) to a phase with completely uncorrelated
heigths (the nematic disorder in the
orthogonal direction to the layer).

We conclude this section noting that if $Z_{[1]}= Z_{[2]}$,
namely if the small $\Delta$ class extinguishes
all the available chain conformations, we have no
transitions at all (of course, since the orientational
order is preserved in our analysis).

\vskip 3truecm
{\bf 4. CHAIN LENGTH DEPENDENCE}
\vskip 1.5truecm

In this section we study the solution of eq. (9)
in function of chain length $N$.
The obtained results depend on the
qualitative properties of $f,\omega,\epsilon$
as functions of $N$; more precisely, they are
determined by the fact that such
functions must have the same asymptotic behaviour
for large $N$ but have different scaling law
for finite size effects.

In order to study eq. (9) we first give
the explicit choice for the packing-flexibility
constraint and the associated representation of
$f_{[1]}$ and $f_{[0]}$.

\noindent
We assume that molecules in plaquette state 1 can substain only
{\it trans} chain conformations; furthermore,
we assume that molecules in plaquette state 2 may have
{\it any} conformation.
Consistently, we take:
$Z_{[1]}=1$ and $Z_{[2]}=Z_{Tot}^{(N)}$.
As for $Z_{[0]}$, the
structure of plaquettes, with two link states (1),
i.e. conformationally "free" chains,
and two link states (0), i.e. packed chains, suggests
$Z_{[0]}=(Z_{[1]}Z_{[2]})^{1\over 2}$.

\noindent
(Alternatively, one may describe a system
where at most $\Delta=1$ conformations are compatible
with plaquette of type 1; in this case:
$Z_{[1]}=1+Z_{\Delta=1}^{(N)}$)

The dependence on $N$ of
$f_{[1]}$ and $f_{[0]}$ is completely determined by Flory's
theory. We consider the extremal cases: with $N<4$, we have no
conformational defects and then $f_{[1]}=f_{[0]}=0$;
with large $N$, the function $f_{[1]}$ takes
the asymptotic form:
$$f_{[1]}^{(N)}(\beta)=N f(\beta)+\eta(N,\beta)\eqno (16)$$
\noindent
where $f$ is independent of N and the function $\eta$
represents the finite size correction.
It is rather easy to verify that such correction decreases
{\it exponentially} with $N$ as a consequence of the
(mono)dimensionality of the chain;
for example, one may extimate $f_{[1]}^{(N)}(\beta)$ with
transfer matrix method.
Analogous considerations apply to $f_{[0]}$.

The problem is to model the dependence on $N$ of
the energy parameters.

\noindent
Contrary to the previous functions,
such quantities are strongly affected by finite size effects
and we are just interested in a range of lengths
($N \sim 10 \div 40$)
where such effects have to be considered.
In order to justify our choices in modelling, we stress here
some points.

\noindent
Let us consider the Lennard Jones interaction between two
molecular chains.
With faced and straight chains, as in the link state ($0$),
the interaction energy behaves asymptotically like
$N \times  atan(N)$; this implies linearity in $N$,
with power law corrections $(-{1\over N}+\cdots)$ for large $N$.
With shifted and disordered chains, as in the link state ($1$),
the interaction has, instead, a dominant term which is
independent of $N$, again with power law corrections.

\noindent
Now consider the dipole-dipole interaction between
molecular heads: the interaction energy is
independent of $N$ in the link state
($1$) (faced dipoles) and is high order
in $1\over N$ in the link
state ($0$) ($N$ measures the dipole-dipole distance).

\noindent
Given a plaquette configuration,
the contribution of next nearest molecules
in the plaquette is essential
in determinig the {\it sign}
of the total first correction in $1 \over N$
to the plaquette energy.
For example, in the plaquette configuration $2$, the next
nearest neighbours chains are faced and statistically closer
than in the plaquette configuration $1$;
this fact may change the sign of the $1 \over N$ term
in the sum which determines $\omega  $.

\noindent
Carring on such an analysis, we obtain the following
general form:
$$\omega= N \omega_0(1+\Omega(N))\eqno (17)$$
and the analogous for $\epsilon$.
In this expression, $\Omega(N)$ is the finite size correction,
with power law dependence on $1\over N$.
In particular, for what said above,
the {\it sign} of the first correction is
characteristic within a given system.

We now search the solution of eq.9 in the form
$\beta^*(N)=\beta_\infty + \delta \beta(N)$ ; we obtain
two equations. The first one is independent of $N$:
$$ -\omega_0 =f(\beta_\infty) \eqno (18)$$
the solution $\beta_\infty $ fixes the order
of critical temperatures within a given class of
homologues; in fact, the quantity
$T_\infty = {{1} \over {k_B\beta_\infty}}$
is the observed value of transition temperature
for large $N$.

\noindent
The second equation, regarding finite size corrections, is:
$$ \delta \beta(N) = {{-\omega_0} \over
{{ {\partial f} \over {\partial \beta}}|_{\beta_\infty} } }
\Omega(N) \eqno (19)$$

\noindent
Since the prefactor is negative,
{\it critical temperatures may increase or decrease
with $N$, depending on the sign of $\Omega(N)$}.

The comparison between
$\beta_C= {{\ln 2} \over {N(\epsilon_0+\omega_0)}}$,
and $\beta^*(N)$ leads to determine one {\it critical
length} $N^*$: we have  $\beta_C(N^*)=\beta^*(N^*)$.

\noindent
According to our results,
for homologues with $N>N^*$ the first order
thermotropic transition occurs, whose main property is
to preserve the layered structure as it
happens in $S-S$ transitions; for
$N<N^*$ homologues, instead, the second order
transition occurs, associated with surface
disgregation as it happens for a smectic layer
at the $S-N$ transition.

These general results agree with observations of
phase transitions in liquid crystals; as said
in the introduction,
many data are available about homologous series of one-chain
molecules, with typical regularities (see for insatnce ref.10).
In particular, a phenomenological expression, similar to the
one used for phospholipids and bilayers $^{(47,48)}$,
describes the observed S-S transition temperatures
as a function of the number of Carbons per chain.
We stress that both the characteristic polynomial dependence on
chain length and the infinite chain transition temperature,
which appear in such expressions, in our case come directly
from eqs. 18 and 19 which give the solution
$\beta^* $ of the model.

\noindent
As an example, we report, in Fig.5, the fitting
of critical temperatures of S-S transition in
alkyl esters$^{(49)}$
and alogeno terminated compounds with alkyl chain $^{(50)}$; in
the figure caption, we report the parameters of the fitting
for each series of homologues.
\vskip 3 truecm
{\bf 5. CONCLUSIONS}
\vskip 1.5truecm
The model presented in this work is a simple, exactly
soluble model of surface which takes into account the flexibility of
the constituent molecules.

It proves the existence of transitions with different character and
order, controlled by the length $N$ of the flexible
section of the molecule (a single linear chain).

\noindent
The character of the transition is determined by the dominant term
in the $1 \over N$ expansion of the interaction energy;
the dependence of critical temperatures on N is instead
determined by the non-leading terms in the same expansion.

The predicted transition leads, with increasing
temperature, from a layered phase (smectic)
to a different smectic or to a nematic phase,
through conservation or, respectively,
disgregation of the surface.

\noindent
In particular, if the surface is  maintained, the
transition is first order; it is associated with
the conformational collapse
of molecular tails and with a change in the mean orientation of
molecules with respect to the surface plane.

\noindent
While at low temperature the surface
is characterized by
exposed molecular heads, which are polar and hydrophylic,
in the high temperature phase the surface is formed by
molecules with aligned heads and exposed, liquid-like
chains, which are hydrophobic and weakly reactive.
Regarding 3D interactions, the surface has a "soft"
character; e.g., we expect a temperature dependent
solubility: in the high temperature regime
the solubility process has an higher energetic cost,
associated with the loss of surface polarity.

As a conclusion we point out two lines of developement
of the present work.

The first regards the reciprocal orientation of nearest
neighbouring head dipoles.
In the present
model we explicitly require  that adjacent head dipoles are opposite;
this defines sites of type $A$ and $B$ in the lattice.
If the requirement is removed, nearest molecules may get
configurations with equally oriented head dipoles.
The work in ref.47 points out the relevant role
played by reversing molecu\-lar heads in liquid crystal transitions.
For our molecules, with polar heads,
the phases in which such reversed
configurations are favoured must exhibit macroscopic polarity.
We think that, at low temperature,
the energetic cost of dipole reversing
is not compensated by the entropic gain, whereas, with increasing
temperature, reversed configurations may be favoured.

\noindent
{}From a technical point of view, to realize this improvement,
the model should be changed by introducing a new link state and a
new energy parameter. In such a situation the analytical treatment
of the model will get much more involved.

The second line regards a more accurate description of tails
interaction. We think it would  be interesting to
consider the multiplicity of minima (decreasing in depth)
which corresponds to relative shifts of adjacent chains of
multiples of the intrachain periodicity (monomer-monomer distance).
Clearly, the link state associated with each one of these minima
represents two partially facing chains, having the {\it active}
section of length $l<N$. If we consider only one of such further
minima, we can  repeat the present analysis noting that,
in this case, we have
$f_{[1]} \propto l$ and $\omega \propto l(1+\omega (N))$.
In so doing we find that the latent heat is proportional to $l$
and independent of $N$, while the behaviour of critical
temperatures is again determined by $1 \over N$ expansion.
These results agree with measured latent heats in S-S transitions
of liquid crystals $^{(52)}$ which are rather constant,
regardless of chain length. We could think
that the interaction among distinct surfaces determines
a mean separation and, as a consequence, favours a specific value
for the active length $l$.

\noindent
We find technically difficult to achieve the solution of a
model with all {\it simultaneously active} minima; on the
other hand, the experimental evidence of a variety of
rehentrant phases and the analysis of such phenomena carried on
in ref.51, suggest the complexity of the problem.
In this regard, we note that if, in our model, we considered the
oscillations around the minima, we would not obtain
a real modification of the present description;
with harmonic oscillations, e.g.,
we could exactly sum over the associated
degrees of freedom and obtain simply a redefinition of the
energy  parameters.
\vfill\eject
\nopagenumbers
{}~

{\bf Figure captions}

Fig.1a

The link state (0) (left) is a local minimum for the attractive
interaction between molecular tails. The link state (1)
(right) is a local minimum for the dipole-dipole interaction
between molecular heads.

Fig.1b

The exchange from link state (0) to (1) is obtained by shifting
an A site molecule downward or by shifting a B site
molecule upward.

Fig.2

Allowed plaquette states and  associated link states:
$(0)\equiv $ full line ; $(1) \equiv $ dotted line.
Correspondence with vertices in the 8V model and related
numbering, according to ref.(38).

Fig.3

Numerical calculation by transfer matrix of the order
parameter as a function of temperature
for different lattice sizes. Dotted line represents the exact
behaviour (see text); in our calculation
$A=\pi ^2 T_C /16 \sqrt 2$.

Fig.4

a - Numerical calculation of susceptibility
as a function of temperature
from transfer matrix results for different lattice sizes.

b - Size dependence of susceptibility at $T_C$ in agreement with
$\chi _L(T_C) \sim L$ (dotted line).
\vfill \eject

Fig.5

Fitting of critical temperatures as a function of the number of
Carbon atoms in the alkyl chain. The fitting has been obtained with:
$\beta^*=1.47993-{1.34839 \over {N}}
+{4.50493\over {N^2}}-{5.90055\over {N^3}}$
 for observed$^{(49)}$ S-S transition temperatures of alkyl
esters ($\bullet $); $\beta^*=1.14883-{{0.08027} \over {N}}$
and
$\beta^*=1.12009-{{0.111465} \over {N}}$
for observed$^{(50)}$ S-S transition of alogeno terminated
aminoketone compounds ($\circ$~-Br and $\overline\sqcup $~-J).
\vfill
\eject
{\bf References}

\smallskip
\item{[1]~~~}
Volkenstein M.V., Configurational Statistics of Polymeric Chains
(Interscience, New York, 1963)

\smallskip
\item{[2]~~~}
Flory P., Statistical Mechanics of Chain Molecules
(Interscience, New York, 1969)

\smallskip
\item{[3]~~~}
Friedel G., {\it Ann.Phys.} {\bf 18} (1922) 273

\smallskip
\item{[4]~~~}
De Gennes P.G., "The Physics of Liquid Crystals" W.Marshall, D.H.
Wilkinson Eds., (Clarendon Press, Oxford, 1974)

\smallskip
\item{[5]~~~}
Gray G.W., Molecular Structure and the Properties
of Liquid Crystals (Academic press, New York, 1962)

\smallskip
\item{[6]~~~}
Gray G.W. and Winsor P.A.  Liquid Crystals and Plastic Crystals
(Horwood, Chicester, England 1974)

\smallskip
\item{[7]~~~}
Demus D., {\it Liquid Crystals} {\bf 5},n.1 (1989) 75

\smallskip
\item{[8]~~~}
Sackman M., {\it Liquid Crystals} {\bf 5},n.1 (1989) 43

\smallskip
\item{[9]~~~}
Adamczyk A., {\it Mol.Cryst.Liq.Cryst.} {\bf 249} (1994) 75

\smallskip
\item{[10]~~~}
Gray G. W., "The Molecular Physics of Liquid Crystals",
G.R. Luckhurst, G.W. Gray Eds. (Academic Press, New York,1979)

\smallskip
\item{[11]~~}
Leadbetter A.J., Richardson R.M., Colling C.N.,
{\it J.de Physique} C1 {\bf 36} (1975) C1-37

\smallskip
\item{[12]~~}
Leadbetter A.J., Frost J.C., Gaughan J.P., Gray G.W. and Hosley A.,
{\it J.de Physique} {\bf 40} (1979) 375

\smallskip
\item{[13]~~}
Levelut A.M., Tarento R.J., Hardouin F., Achard M.F. and Sigaud G.,
{\it Phys.Rev.A} {\bf 24} (1981) 163

\smallskip
\item{[14]~~}
Dabrowsky R., Czuprynski K. et al.
{\it Mol.Cryst.Liq.Cryst.} {\bf 249} (1994) 51

\smallskip
\item{[15]~~}
Frenkel D. "Statistical Mechanics of Liquid Crystals",
Course 9, Les Houches LI 1989,
Hansen J.P., Levesque D. and Zinn-Justine Eds. (Elsevier 1991)

\smallskip
\item{[16]~~}
The original article by Onsager L.,
{\it Ann. N.Y.Acad.Sci.} {\bf 51}  (1949) 62
is not available to us

\smallskip
\item{[17]~~}
see Isihara A. {\it J.Chem Phys} {\bf 19} (1951) 1142 and
Wadati H., Isihara A. {\it Mol. Cryst.Liq.Cryst.} {\bf 17}
(1792) 95 ; see also ref.4 Ch.2

\smallskip
\item{[18]~~}
Mayer W., Saupe A., {\it Z.Naturforsch} {\bf 13a} (1958) 564; {\bf 14a}
(1959) 882; {\bf 15a} (1960) 287

\smallskip
\item{[19]~~}
De Gennes P.G.,{\it Mol.Cryst.and Liq.Cryst.} {\bf 12} (1971) 193

\smallskip
\item{[20]~~}
Mc Millan W.L., {\it Phys.Rev. A} {\bf 4} (1971) 1238;

\smallskip
\item{[21]~~}
Mc Millan W.L., {\it Phys.Rev. A} {\bf 6} (1972) 936;

\smallskip
\item{[22]~~}
Mc Millan W.L., {\it Phys.Rev. A} {\bf 7} (1973) 1673;

\smallskip
\item{[23]~~}
Marcelja S., {\it J.Chem.Phys} {\bf 60} (1974) 3599

\smallskip
\item{[24]~~}
Pink D.A.,  {\it J.Chem.Phys} {\bf 65} (1975) 2533
see also G.R. Luckhurst, in ref.10. Ch.4

\smallskip
\item{[25]~~}
Martire D.E., {\it Mol.Cryst.Liq.Cryst.} {\bf 28} (1973) 63

\smallskip
\item{[26]~~}
Luckhurst G.R., Heaton N.J., {\it Molecular Physics}
{\bf 66} n.1 (1989) 65

\smallskip
\item{[27]~~}
Lebwohl P. A., Lasher G., {\it Phys.Rev. A} {\bf 6} (1972) 426

\smallskip
\item{[28]~~}
Chiccoli C., Pasini P., Zannoni C., {\it Physica A} {\bf 148} (1988)  298

\smallskip
\item{[29]~~}
Kunz H. Zumbach G., {\it Phys.Lett.} B {\bf 257} n.3-4 (1991) 299

\smallskip
\item{[30]~~}
Ferrenberg A.M., Swendsen R.,
{\it Phys.Rev.Lett.} {\bf 61} (1988) 2635;
{\it Phys.Rev.Lett.} {\bf 63} (1989) 1195

\smallskip
\item{[31]~~}
Nagle J.F,. {\it J.Chem.Phys.} {\bf 58} (1973) 252

\smallskip
\item{[32]~~}
Mar\~celia S., {\it Biochimica et Biophysica Acta} {\bf 367} (1974) 165

\smallskip
\item{[33]~~}
Abraham D.B.,
"Surface structures and phase transitions - Exact Results"
in Phase Transitions and Critical Phenomena,
C.Domb., J.L.Lebowitz Eds. (Academic Press, London, 1986)
{\bf 10} 1

\smallskip
\item{[34]~~}
Van Beijeren H., {\it Phys.Rev.Lett.} {\bf 38} (1977) 939;

\smallskip
\item{[35]~~}
Van Beijeren H. and Nolden I. "Topics in Current Physics",
W.Schommers and P.von Blanckenhagen Eds.
(Springer-Verlag Berlin 1987) {\bf 43} Ch.7 259

\smallskip
\item{[36]~~}
Burton W.K., Cabrera N., Frank F.C.,
{\it Phyl.Trans.R.Soc.} {\bf A243} (1951) 299

\smallskip
\item{[37]~~}
Lieb E.H.,
{\it Phys.Rev.Lett.} {\bf 18} (1967) 1046

\smallskip
\item{[38]~~}
Baxter R.J., "Exactly Solved Models in Statistical Mechanics"
Academic Press, London, (1982)

\smallskip
\item{[39]~~}
Baxter R.J., {\it J.Phys.C: Solid State} {\bf 6} (1973) L94

\smallskip
\item{[40]~~}
Mazzeo G.,Jug G., Levi C.A. and Tosatti E.,
{\it  J.Phys.A:Mat.Gen.} {\bf 25} (1992) L967-L973

\smallskip
\item{[41]~~}
Mazzeo G., Jug G., Levi A.C., Tosatti E., {\it Surface Science}
{\bf 273} (1992) 237

\smallskip
\item{[42]~~}
M.Kikuci, K.Binder, {\it Europhysics Letters} {\bf 21} (4) (1993) 427

\smallskip
\item{[43]~~}
Zerbi G., "New Perspectives of Vibrational Spectroscopy
in Material Science, Advances in Applied F.T.I.R.
Spectroscopy, Mc Kenzie Ed. (Wiley, New York 1988)

\smallskip
\item{[44]~~}
Leibler S. in "Statistical Mechanics of Membranes and Surfaces"
Vol.5, Nelson D.,Piran T and Weinberg S. Eds.(World Scientific 1989)

\smallskip
\item{[45]~~}
Mazzoletti G. Thesis, Universit\`a degli Studi di Milano (1993)

\smallskip
\item{[46]~~}
Bassetti B.,Benza V.,Jona P., {\it J.Phys.France} {\bf 51} (1990) 259

\smallskip
\item{[47]~~}
Cevc G., {\it Biochemistry} {\bf 30} (1991) 7186

\smallskip
\item{[48]~~}
Huang C., {\it Biochemistry} {\bf 30} (1991) 26

\smallskip
\item{[49]~~}
Gray G.W. see ref.9, Ch.12, fig.2

\smallskip
\item{[50]~~~}
Pi\'zuk W.,Kr\'owczynski A. and G\'orecka E.,
{\it Mol.Cryst.Liq.Cryst.} {\bf 237} (1993) 75

\smallskip
\item{[51]~~}
Netz R.R. and Berker A.N. "Phase transitions in Liquid Crystals",
Martellucci S. and Chester A.N. Eds. (Plenum, New York 1992)
Ch.7

\smallskip
\item{[52]~~~}
Coates D.,Harrison K.J. and Gray G.W.,
{\it Molecular Crystals and Liquid Crystals} {\bf 22} (1973) 99
\vfill\eject
\bye